# Do animals share space? An explanation to Zeno's paradox (of motion)[1]

Manuel A. B. Bache* · Jorge Hernández Contreras**

*HUM-962, University of Sevilla. ORCiD 0000-0002-7606-5252 ; mbejarano@us.es

**University of Guadalajara. ORCiD 0000-0001-5290-7307

## Abstract

In this paper, a solution to Zeno's paradox of motion was offered and a possible, based on observations and empirical viewpoint, interpretation. A *gedanken* experiment was proposed in accordance to our empirical and experimental paradigm, offering solutions from pendular energy and geometric gravity theory, the standard view, and classical physics, as a reply to Zeno's paradox (of motion), departing from the logical and overly mathematical frameworks, which either offered a solution detached from experiment and observations, the former, and a solution with neither explanation nor interpretation, the latter.

It takes on the problem of unmeasurable regions, Zeno's paradoxes (measure and motion), and infinitesimals, offering some insights that could be found interesting for both, metaphysical and foundations of physics fields.

---

[1] Special thanks to the referees of *Humanities and Social Science Communications* for their very helpful comments. Acknowledgments to classical Boltzmann, Newton and Leibniz, and to Witten, Einstein, and Planck, broad shoulders onto which hold on. I'd also like to thank *The Portal* participants, especially Eric Weinstein and Roger Penrose. Special appreciation to Neil Turok's, Poincaré's and Penrose's works.

# 1 Introduction

Since the age of classical Greek philosophy, some paradoxes have been arising regarding physics and motion. On the one hand, one may refer to wave-particle duality (Huygens, 1690; Newton, 1704), which even firstly swayed to the second due to Newton's prominence and Poisson's positivist perspective, it was finally proven and set that paradox (see *Poisson spot*: Arago & Fresnel, 1818; Poisson, 1818); to the infamous Einstein-Podolsky-Rosen (EPR) paradox (Einstein, Podolsky, & Rosen, 1935); or to the irreconcilability of Einstein (1915; 1916) general relativity (GR) and quantum mechanics (QM) theory (Aspect, Dalibard, & Rogers, 1982; Bell, 1976; Bell et al., 1985; Davisson & Germer, 1927; de Broglie, 1923; 1924; Einstein, 1905; Heisenberg, 1927; Planck, 1901ab; Schrödinger, 1926; 1935a; 1935b; Wheeler, 1978; Young, 1804; Zeilinger, 1986), irreconcilability that has been already pointed out (Deser, Tsao, & Van Nieuwenhuizen, 1974; 't Hooft, 2005, p. 29; 't Hooft & Veltman, 1972; 1974; Nishino & Rajpoot, 2008; von Borzeszkowski, Treder, & Yourgrau, 1979; Weinberg, 1972, p. 289).

On the other hand, one may highlight, perhaps less known paradoxes, such as Mpemba paradox (regarding non-linear behavior to freezing-cooling events and the freezing process: Jeng, 2006; Kumar & Bechhoefer, 2020; Mpemba & Osborne, 1969; Tao et al., 2017), Ehrenfest paradox (Ehrenfest, 1909; Grøn, 1979; Kumar, 2024), or, distinguishing it in this paper, Zeno's paradoxes, especially Zeno's paradox of motion regarding Achilles and the continuum problem that was tackled by his master Parmenides (see Chen, 2021; Gale, 1968; Grünbaum, 1968; 1973, to expand into the specific paradox).

## 2 Zeno's Paradox Solutions and Their Problems

Zeno's paradoxes are a collection of at least 9 paradoxes targeting metaphysical problems in order to back his master's (Parmenides') philosophical doctrine regarding the spatial continuum (a pervasive-like universal monism) and the illusion of motion (immobilism), rejecting dichotomy and the existence of individual self; that can be found within Aristotle's work on physics, wherein Aristotle discussed them and offered solutions to some of them. One of the most famous Zeno's paradoxes is the so-called "Achilles and the tortoise" (AR, standing for Achilles race). Achilles footrace paradox (AR) is established on some assumptions like Infinite Divisibility; Infinity Conditional; Dichotomy; Additivity (*vide* Chen, 2021, p. 4442) and the *prima facie* unconditional defense of Parmenides' metaphysical arguments (self-confirmation bias).

There have been some attempts to solve AR paradox (Cauchy, 1821; Lynds, 2003; Mayila, Mpimbo, & Rugeihyamu, 2024; Russell, 1914; van Bendegem, 1987; Weyl, 1949), even from the quantic point of view (Schlegel, 1948), remaining an open metaphysical problem and question (Chen, 2021; Gale, 1968; Mayila et al., 2024; Papa-Grimaldi, 1996). One may refer to Cauchy (1821, p. 124) for an accurately mathematical solution, and Grünbaum (1973, pp. 158-176) for the standard solution based on standard analytic geometry. As Chen (2021, p. 4443) stands out, according to this view, "a geometric line can be algebraically represented by the set of real numbers, with each real number representing a point on the line", meaning that a line can be considered exhaustively composed of points. However, the standard view violates additivity, inasmuch as it assumes that a line is composed by points of zero length, and an infinite number of points of zero length add to zero:

$$\lim_{n \to \infty} (n \cdot 0L^1) = 0$$

□

Even though from the point of view of the foundations of physics and particle physics, it could be considered seemingly correct, it makes an argument in regard to the quantum foam (Wheeler, 1955; Wheeler & Ford, 2000), and the concept of mass (*vide* Haug, 2023) for the special case within the QM framework (*vide* "sea-quark": Andersson, Gustafson, & Peterson, 1977; Brodsky et al., 1980; van Hove, 1987; and quarks: Bjørken & Glashow, 1964; Gell-Mann, 1964; Zweig, 1964).

With that in mind, regarding the problem of additivity, insofar π is considered an irrational number, one may propose a proof for a closed line considering infinite divisibility locally:

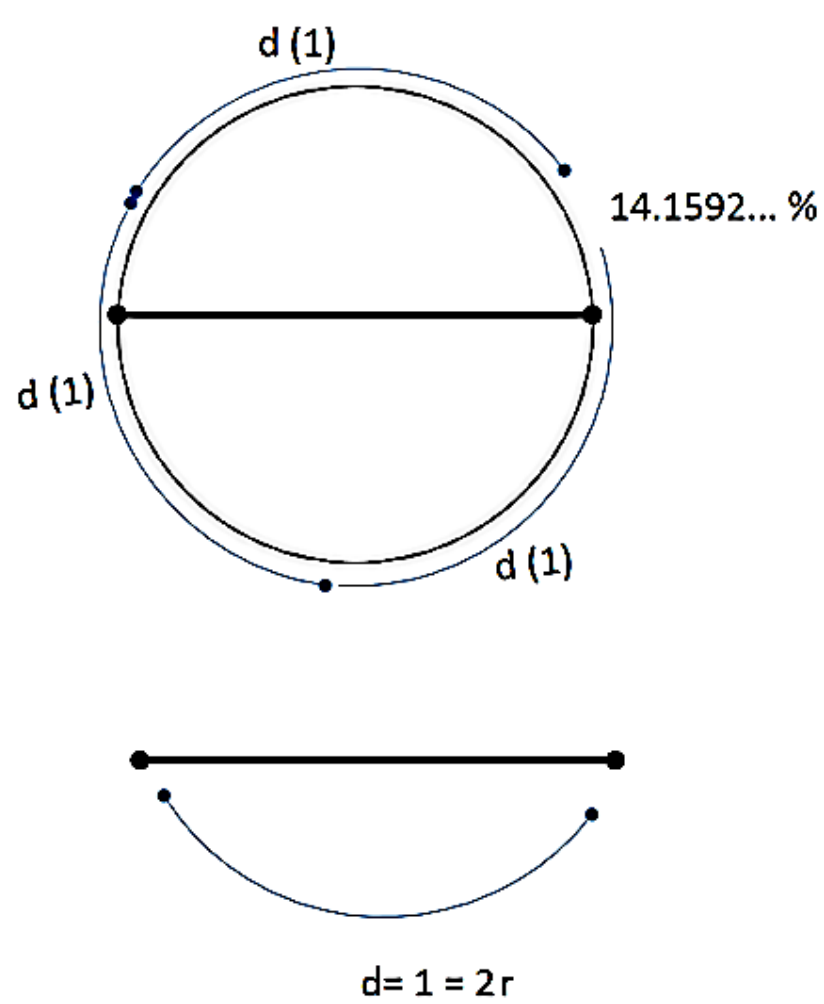

**Figure 1**

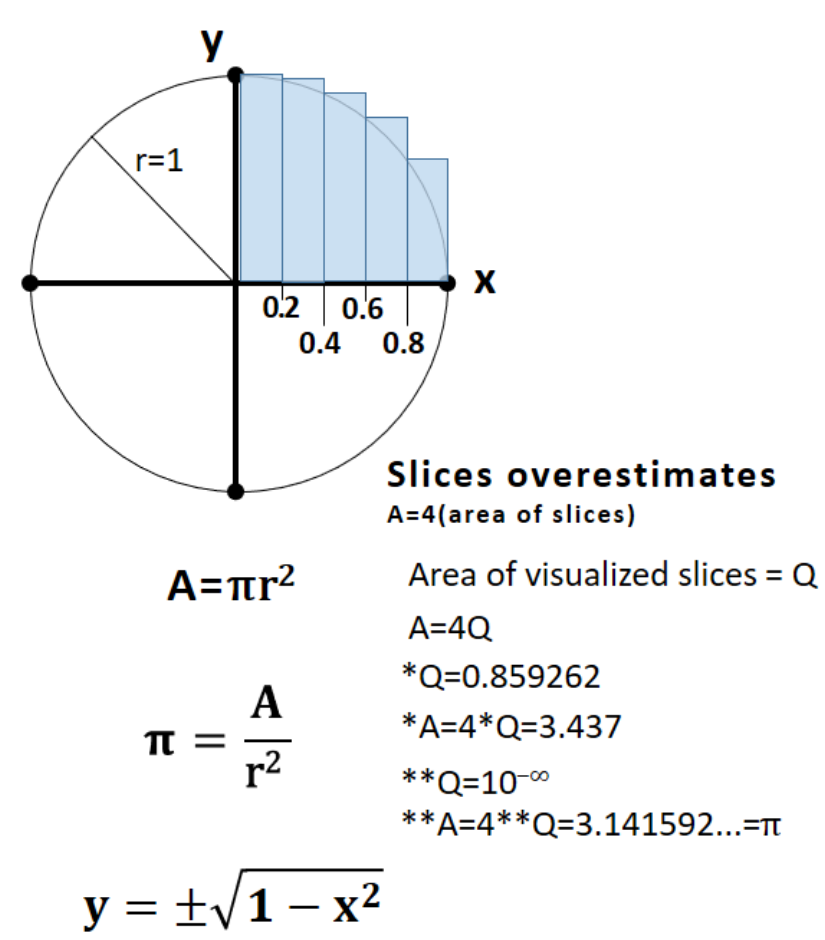

**Figure 2**

In Figure 1 diameter was set as 1 to offer an intuition about the relation between diameter *d* (and, hence, radius as well), and the circumference's length ($L_C$). In Figure 2 radius ($r$) was set as 1, and we explore the integration of the area of the 4 quadrants (Q) by the integration of slices (first splicing the radius in 4: 0.2, 0.4, 0.6, 0.8; 5 slices; then reducing y, i.e. the slices' width (x), to $10^{-\infty}$), so one can obtain that $A = 4 \cdot Q^{**} = 3.141592\ldots = \pi$, an exacting solution to π, with exacting accuracy. It hints that even the width of the lines may be important for a solution.

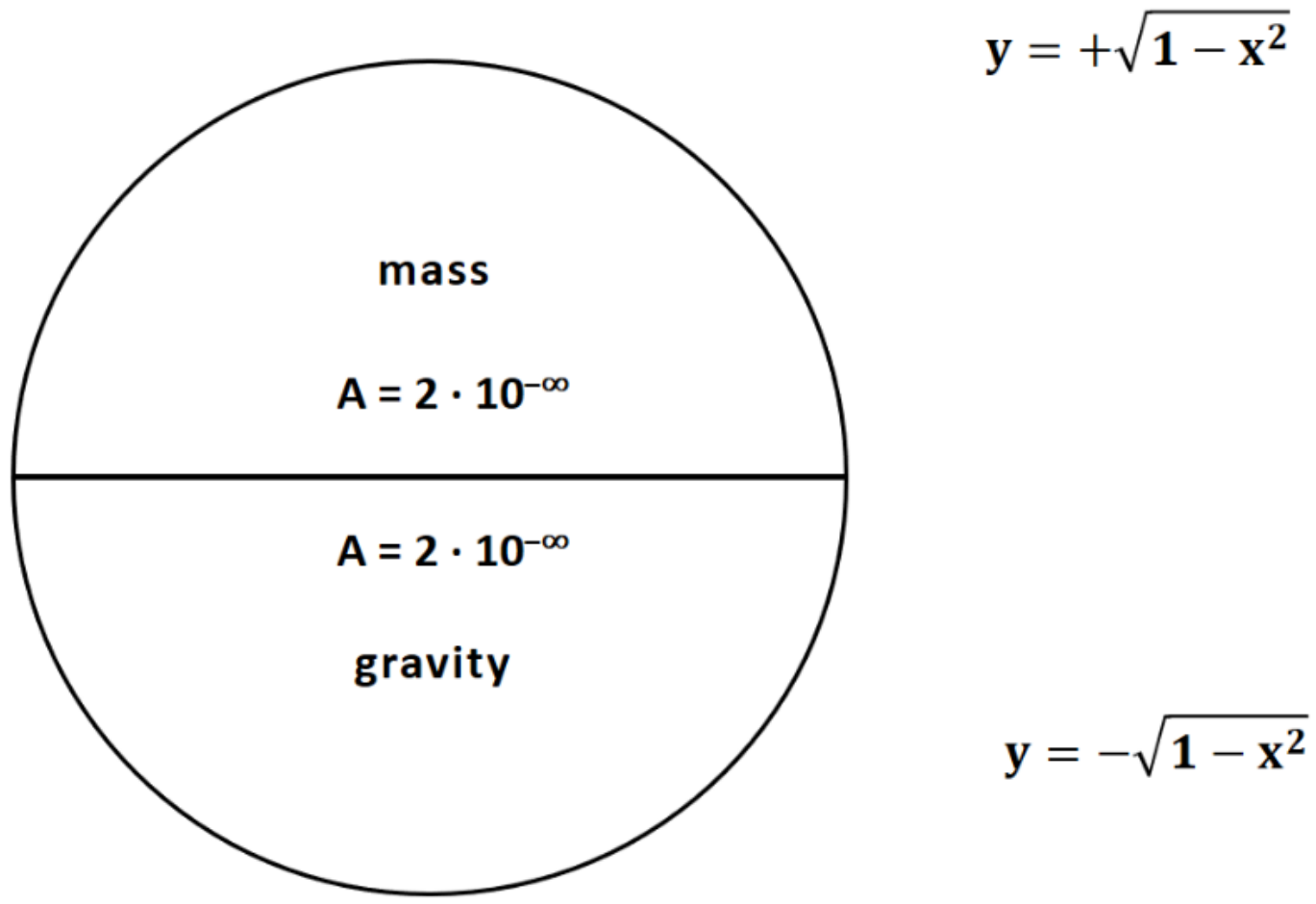

**Figure 3**

In Figure 3 one can establish a relation between mass and gravity in accordance to the equation for a circle $x^2 + y^2 = r^2$, as long as we set $r = 1$ , then $y^2 = 1 - x^2$; thus $y = \pm\sqrt{1 - x^2}$ , where one may consider the positive part $y = +\sqrt{1 - x^2}$ as mass, and the negative part or direction (*vide* Dirac, 1931; 1975) $y = -\sqrt{1 - x^2}$ as gravity.

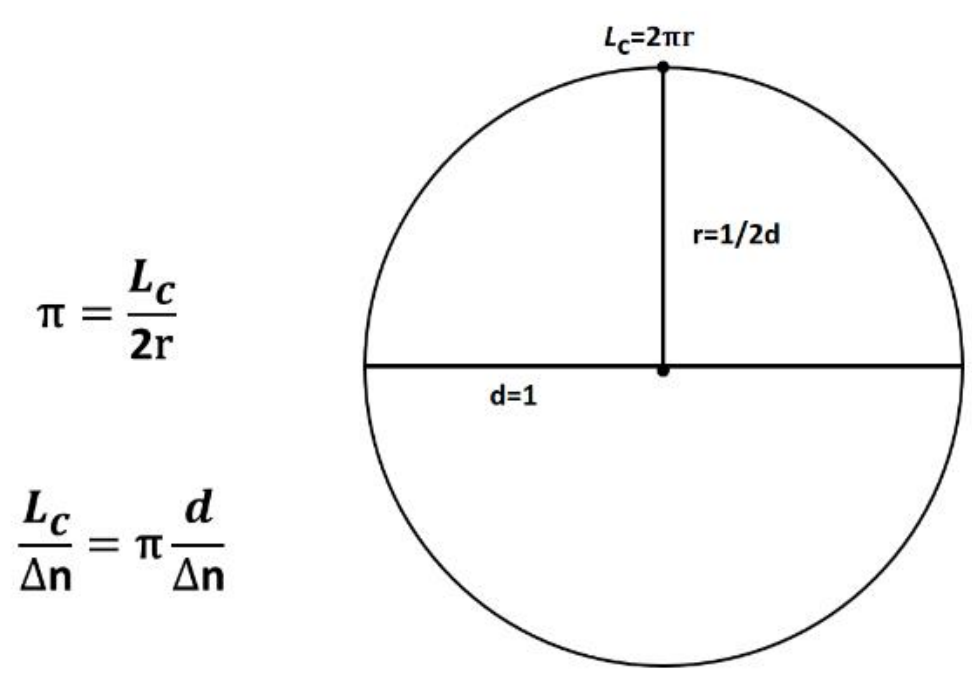

a)

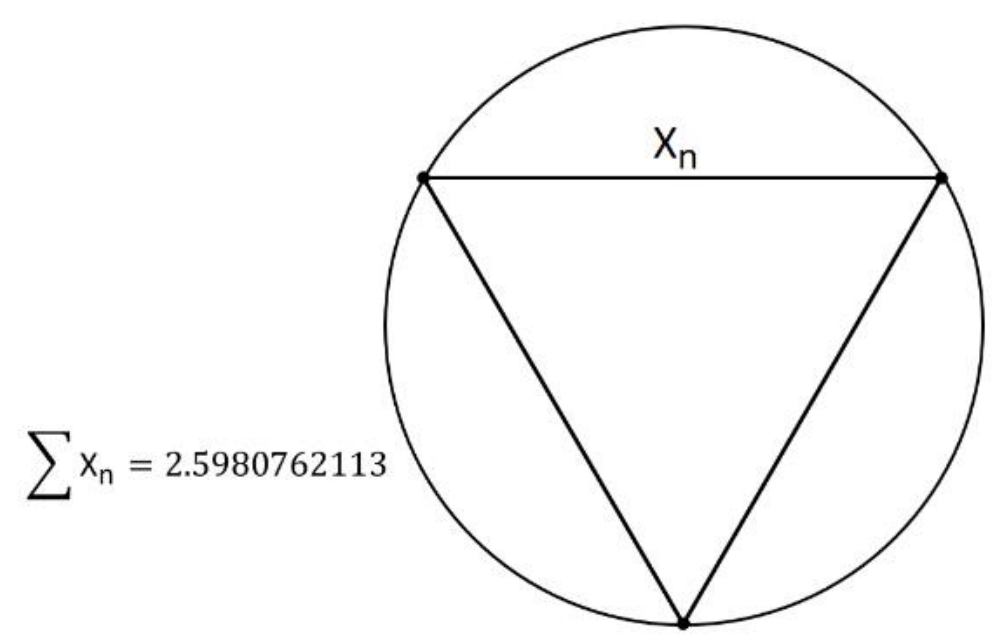

b)

**Figure 4a)** **Figure 4b)**

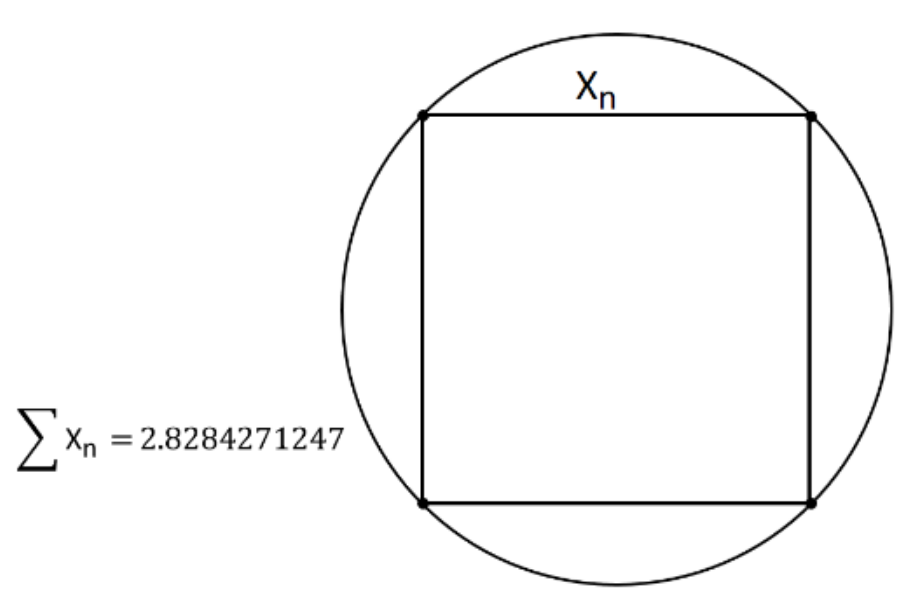

a)

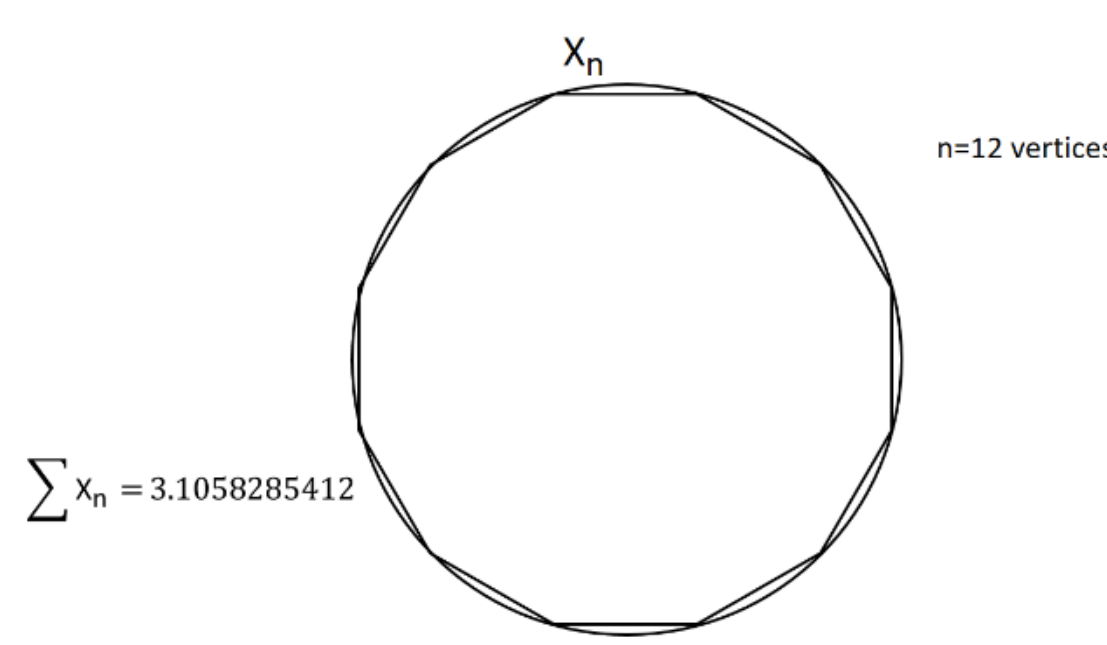

b)

**Figure 5a)** **Figure 5b)**

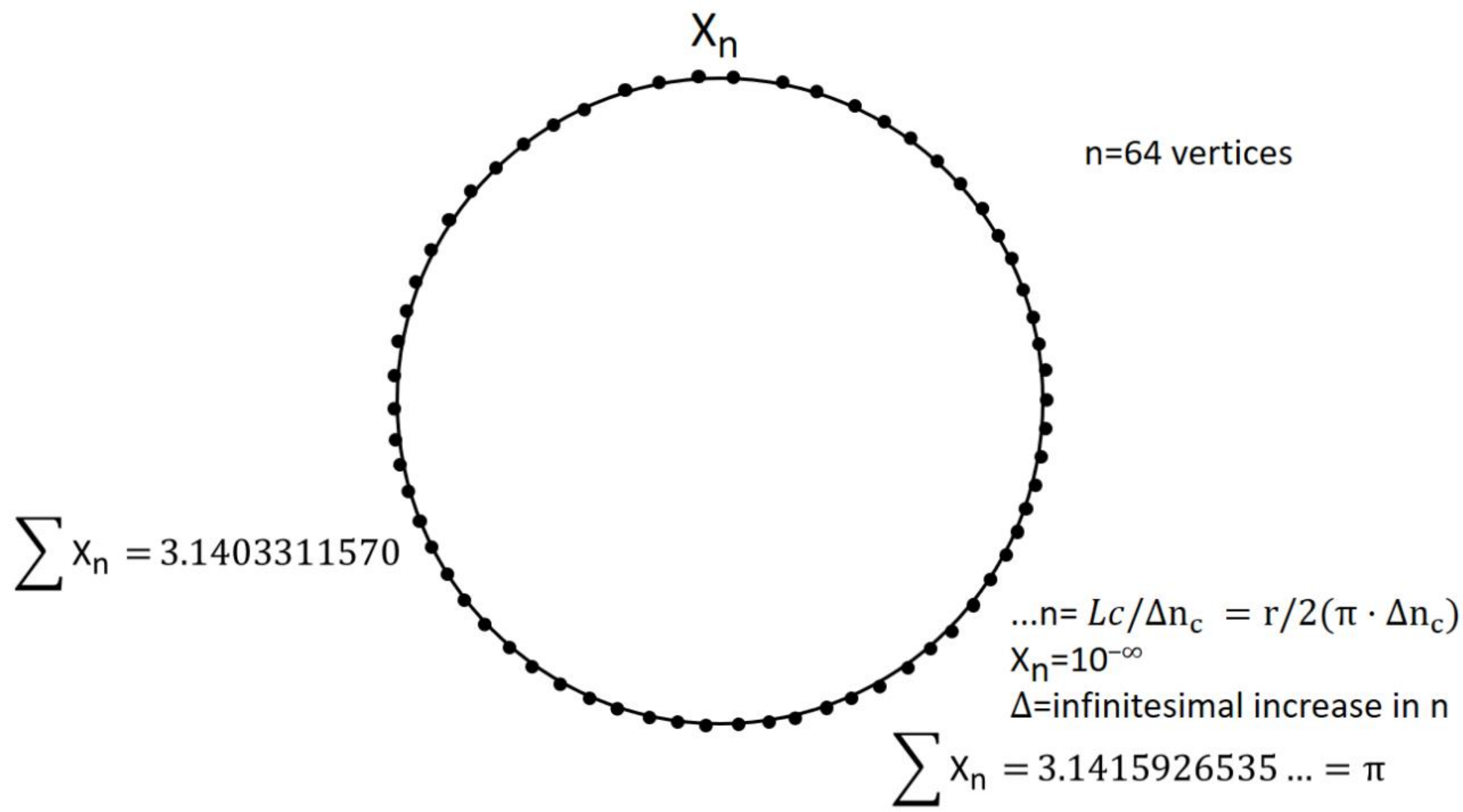

**Figure 6**

In Fig. 4a) one may observe a relation between π as the length of the circumference ($L_C$) divided over 2 times the radius (r), and the increase on both sides number of points (n), as the increase Δn of the number of points when one splits the circumference in more points will yield a proportional increase of n in diameter and/or radius; and one can obtain a first approximation to π by inscribed n-vertices polygons, in this case a triangle (3 sides and 3 points). In Figure 5a) and 5b) one may observe that the accuracy tends to increase as the number of vertices (n points) or sides of the inscribed polygon increases, from 2.5980762113 for a triangle in Figure 4b) to 2.8284271247... for a square, and 3.1058285412 for a dodecahedron in Figs. 5a) and 5b) respectively. Finally, one can realize the ratio between the point-to-point distance ($X_n$) and the accuracy as the inscribed polygon's sides and points (vertices) increase, as the $\sum(X_n) = L_c$ in Fig. 6, and once $X_n$ was set as $10^{-\infty}$ , $\sum(X_n) = L_c = \pi$. Thus, let $X_n$ be set as $10^{-\infty}$ the distance between circumference's points will be proportional to 2 times π times the distance between radius' points, or π times the distance between diameter's points, and π represents the ratio between the points, insofar a new point is added to n.

□

One can sit (and wait) on the non-denumerability of π and of the countless infinitely divisible n points, to add up to the proof.

Perhaps, the best way *to prove a point* is trying to solve the "*before the beginning problem*" for the universe and the Big Bang (Bache, 2023a; 2024a, pp. 119-124; Hawking, 1974; Penrose, 2010; 2020), which will be tackled in the next section.

For the problem on additivity and on solutions to AR paradox, one may refer to the aforementioned works and a further review to Chen (2021).

# 3 Infinitesimal Atomism

For the influential ideas of Greek philosophy and the counterintuitive yet proven physical properties of QM (Aspect et al., 1982; 2022; Bell et al., 1985; Einstein, 1905; Zeilinger, 1986), one may introduce this section, devoted to a problem (a related problem to the purpose of this paper) referred as "logical atomism" (see Democritus, 450 B.C.E., and Russell, 1924), as well as to an effort to review research on physics, even though the math sometimes can be harsh, and sometimes the understanding of them may seem rough, just to take a sense on the inconsistencies and contradictions, as well as avoiding to miss experimental and empirical research to track their own research. Especially, one may highlight Prigogine's works (Prigogine, 1977; 1989; Prigogine & Nicolis, 1967; Prigogine & Stengers, 1984) on emergence, as an ambiguous property that even physicists sometimes do not reach to understand; and the concept of entropy, by pointing to works like Leff (1996), Sharp & Matschinsky (2015), and Styer (2000), as a physical concept and property, that even most physicists working in the field do not understand (attributable to von Neumann's words and 1927 work, and Shannon, 1948).

While fermions (matter) can neither be regarded as infinitely divisible, nor as negative, from the QM point of view, keeping it within standard analysis (SA), i.e., its length will be kept into the real line R, on the other hand, bosons (forces) can be regarded as infinitely divisible within the point of view of QM (Bose, 1924; Hanada, Shimada, & Wintergerst, 2021; Sen & Gangopadhyay, 2024; Shieh, 2015), i.e. its analysis can be extended to non-standard analysis (NSA) in such a way that its consideration of length (and superposition) can be easily considered and viewed into an extended R line, or hyperreal line *R, fulfilling some of the hardest QM properties.

Regarding the analysis of fermions within the setup posed by Zeno's AR paradox, whether we call them "minims" (Chen, 2021), quarks (Gell-Mann, 1964) or monads (Leibniz, 1720), they can be required to comply to current observations and experimental research. Whether we agree or not, Planck (1901ab) set a limit in energy, as he called it *quanta*. Following that path, although keeping it open to arguments and fruitful debate (Bache, 2023ab; Hamada, 2022) it sets a limit to length ($\ell_p$), at least concerning fermions, and it may also set a limit to length regarding space (Mead, 1964, p. B857). As aforementioned, if one is up to tackle on it, one may try retrying the old problem to the origin of (at least) our universe, described as "while the space–time Ricci curvature is singular, the Weyl curvature is not: he (Penrose) conjectures that it must be finite or zero" (Tod, 2024, p. 6), and: "the metric and Ricci tensor are singular but the Weyl tensor is not" (Penrose, 1979, pp. 630-631; cfr. Tod, 2024, p. 6), that we regard as a flat Minkowski space or Weyl curvature hypothesis (Bache, 2024a, p. 120), and a shared coordinate x' (Bache, 2023ab; 2024a). Thus, the interaction between what one may call a Planck string (of length $\ell_p$) with another Planck string (of length $\ell_p$), it would become a surface, i.e., a flat space of $(\ell_p)^2$ .

Let the origin of the universe be a flat Minkowski space with a shared coordinate x' related to the Weyl curvature hypothesis, its length becomes

$$\ell_{\mathrm{p}} \cdot \ell_{\mathrm{p}} = (\ell_{\mathrm{p}})^2$$

$$(\ell_{\mathrm{p}})^2 = 2.612280225025 \cdot 10^{-70}\ \mathrm{m}^2$$

(a surface), where $\ell_{\mathrm{p}}$ is Planck length, approximate values of $\approx 1.616255 \cdot 10^{-35}$ m.

As a fundamental length $\ell_{\mathrm{p}}$ can be regarded as a point-like or as a one-dimensional Planck string of Planck length ($\ell_{\mathrm{p}}$), and, in the origin (x'), as a fundamental space where even space, mass and fermionic components seems to become fuzzy and intriguingly continuous (Hanada et al., 2021; van Hove, 1987; Wheeler, 1955) it might be $(\ell_{\mathrm{p}})^2 = 2.612280225025 \cdot 10^{-70}\ \mathrm{m}^2$ which no one claims that that needs to be squared (a quadrilateral flat space or regular quadrilateral surface), but, on the contrary, it may well be of curved lengths, corresponding with the Weyl curvature hypothesis.

Not to mention that, set that very space as a one-dimensional string, it hints to a reduction of space placed on a real line R, instead of increasing its length (Bache, 2024b), predicting an inversion of the usually proposed and known as bouncing-cosmology (Barros, Teixeira, & Vernieri, 2020; Popławski, 2012) before the beginning, or a transplanckian epoch (Hamada, 2022), something that, perhaps, may lead and be linked to an accelerated expanding universe, as observed in cosmology (Perlmutter et al., 1998; Riess, Schmidt, & Perlmutter, 2011; Schmidt et al., 1998) .

For the everyday life world and space, the hyperreal extension, and the relation to how humans usually perceive, yet rather a physics discussion added to it, one may recommend Chen (2021, pp. 4444-4446), and the review on the topic that was already discussed by Gibson (1966, pp. 20-27; 1979) regarding primitives of the natural world, and the affordances for human representation and perception of the world.

In regard to the hyperfinitiness, regarding the analogy used in order to discuss it, i.e., the origin of the universe, for brevity one may point to Chen (2021, pp. 4446-4448), Bache (2023ab; 2024, pp. 112-140), Hanada et al. (2021), Krauss (2012), and Weyl (1949) as well, and highlight that, as in the origin the fermionic properties seem to decay towards the Mead's limits fuzziness, and into a fermionic-bosonic fusion (Andersson et al., 1977; Brodsky et al., 1980; Hanada et al., 2021; Krauss, 2012; van Hove, 1987; Wheeler, 1955), attending partially the irreconcilability of GR and QM and the known, usually referred as, *symmetry-breaking problem*, it seems recommendable the rejection of Democritus logical atomism (as pointed out in Bache, 2023a, p. 270, pp. 285-291; 2024a, p. 117, pp. 130-134; Chen, 2021, p. 4444) in favor to dualism and dichotomy (Bache, 2024ab; Boyle, Finn, & Turok, 2018; Penrose, 2010; 2020), also referred in NSA as the hyperfinite collections assumption or unrestricted composition, which assumes that "collections of regions of a special kind—hyperfinite collections of regions—have fusions" (Chen, 2021, p. 4446). And, not necessarily every, but, perhaps, "every collection of regions has a fusion" (Chen, 2021, p. 4446). In order to avoid excessively expanding this section, in regard to the application of the Banach-Tarski paradox (Banach & Tarski, 1924), the distribution within the shell finds its limit in the 5th dimension (Bache, 2024a, p. 135; 2024b; Gardner, Koldobsky, & Schlumprecht, 1999; Guédon, 2014, pp. 48-54; Tao, 2011), so a mixed-state system between 4-5 dimensions might still expand and whenever it reaches the 6th dimension, it would become unstable, probably in a regressive fashion, setting a steady-state system, or the kind (Bache, 2024a, p. 139-145; 2024b), and let us consider that the Schwarzschild singularity may be a collision, or, perhaps, the asymmetric inversion of (fermionic) distribution of two interacting fundamental particles or events (Bache, 2024a; Boyle et al., 2018; Penrose, 2010; 2020), then it may holds for the transplanckian epoch or space as Hamada (2022) proposal and others previous Hamada's works argued; and the Weyl curvature hypothesis as a potential Minkowski flat space; to solve the hyperfinitiness and the unmeasurable regions problem.

# 4 On Non-Measurable (infinitesimal) Regions

Whether the infinitesimal atomism avoids Zeno's AR paradox by dispensing with dichotomy (as we have proposed in the previous section, and also can be further reviewed within the referred works, and Chen, 2021, pp. 4442-4448), or it does not, as it depends on how well it satisfies other intuitive assumptions in the paradox (Chen, 2021, p. 4448), considering divisibility, fermionic matter cannot hold both principles: Finite Divisibility (every finitely extended region can be divided into smaller parts), and General Divisibility (every extended region can be divided into smaller parts). Although fermionic matter seems to fulfil General Divisibility if one assumes van Hove (1987) sea-quarks and Wheeler's quantum foam, known as geons (Wheeler, 1955; Wheeler & Ford, 2000), it would not fulfil Finite Divisibility considering its physical energy version *quanta* (Planck, 1901ab) neither would fulfil it regarding length if assumed Mead's limit for fermionic matter.

> Then, how may one propose a solution to Zeno's AR paradox, as well as an empirical interpretation for it, if, *a priori*, even from the physics and the most fundamental physical properties, one cannot set both principles, required for a solution within the current empirical scientific paradigm (i.e., not only a solution, but also an interpretation within the current experimental framework)?

The key might be in the "*a priori*". Since the current framework is not unified, the geometric solution that offers a viable solution (the standard view, *vide* Chen, 2021, pp. 4443-4444; Grünbaum, 1973) tends to be either rejected or not fully accepted, as long as the trend seems to be the rejection of GR, i.e., the only theory that might offer a viable solution, yet not understood insofar it is not unified with the observations in QM, and it loses its physical meaning within the QM framework.

It is already known that the standard view satisfies both; finite atomism satisfies neither; and infinitesimal atomism satisfies Finite Divisibility but not General Divisibility (Chen, 2021).

Hereby we will grant that every confining mass body within the same (our) gravitational field, is an individual quantic-body. To some extent, it means that their interaction with gravity will be local and independent of other bodies. However, our proposal aims to the understanding of the solution proposed, as we believe that the problem lies on the comprehension, or the lack of understanding, of gravity.

## 5 Solutions and Their Problems. (On the Problem of Non-Measurable Regions)

We will propose a solution from the geometric point of view, since it allows to better tackle into the hyperfinitiness problem and the fusion of hyperfinitely many measurable regions, as they become or may become unmeasurable (immeasurability that would be leading to the acceptation of Zeno's AR paradox, which *gedanken* experiment one can check and falsify in everyday life experiments and situations).

Thus, although Zeno's AR paradox can be refuted just by the implications of Planck (1901ab) works, and some of the proven properties of QM, our purpose in the present work is not (or is not limited to) the rebuttal of Zeno's AR paradox, but, by offering an interpretation to the interwoven event, offer an explanation to both, the solution and the divergence between Zeno's *gedanken* experiment and the most frequently incoming outcome.

We will not, indeed, assume the outcome $\frac{1}{2}+\frac{1}{4}+\frac{1}{6}+\frac{1}{8}+\cdots=1$ , i.e. we restrict our setup to $\frac{1}{2}+\frac{1}{4}+\frac{1}{6}+\frac{1}{8}+\cdots\neq 1$ aiming to (at least) an exacting solution.

Neither will be our result set as a mathematical proposal, nor a geometrical without explanation either. Indeed, one may check that, if we assumed Unrestricted Composition, once the participants are close one to each other, then there are many more (increasingly) unmeasurable regions, so the aforementioned mathematical assumption (series) would lead to an approximate result, yet it also would remain in the Zeno's paradoxical realm.

Thus, let a lynx moving at 17 m/s set at a distance of (e.g.) 22m from a snail that were able to escape at 1.55 m/s (same direction), according to Zeno's AR paradox, the lynx will never catch the snail, due to the inability of the faster animal (the lynx in that very case) to overcome the infinitely splitting of that extending space, showing the asymptotic behavior of the assumption of infinitely divisible space, i.e. the asymptotic behavior of the graph of 1/x. According to that *gedanken* experiment, it always gets closer to the X-axis, but it never reaches 0 (never touches it, their coordinates never match in space, or the lynx will never catch the snail).

Now, one may first offer a solution, and afterwards proceed to unfold the solution, fleshing out the explanation in the next and final section.

The problem may arise from two operators (leading the predictions from biomechanics and classical physics) setting the path that finally will take in nature: one, the reason why one animal moves faster than the other (the main feature that makes faster to the faster animal); two, the assumption of uniformity, i.e., the animal A (the lynx) and B (the snail) being two linear vectors travelling at constant speeds, instead of composite tensors acting and working as non-linear vectors. Both are leading to our answer, and the reason of using GR (and that the standard view offers the standard solution), yet it would not always produce the deterministic course of the solution (remaining within the QM framework). These will guide our answer and the answer that nature tends to give to Zeno's AR paradox, as *beables*.

Then, let a lynx (A) set at a random distance of 22m from a snail (B) travelling both in the same direction, A at 17 m/s and B at 1.55 m/s, A with the goal of catching B, the lynx

(A) will get a snail (B) in $23.55/17 = 1.3852941$; $(1.3852941 - 1)/17 = 0.02266436$

$t = 1.3852941 + 0.02266436 = 1.4079585$ s , [2]

Should this solution be postulated, it can be used for other *gedanken* experiments with animals, and Zeno's AR paradox examples, setting their speeds, and a spatial separation (x), whenever A is assumed to empirically catch B.

For the very same example, A(x) – B(x) = 22m. After 1s, the snail (B) is at 23.55m, and the lynx (A) will be at 17m, at a distance of 23.55 – 17 = 6.55m. It is expected that, within the next second, A will catch B, since, in 1 additional second, the lynx will travel 17m, while the snail will travel 1.55m. Assumed that within the lapse of that very second 2, A will reach B (empirical view over the positivist view), let the lynx (A) be at 6.5m, reaching the snail (B) within the interval [1-2s], both keeping their respective speeds, their distance ($x_{A-B}$) will be 0, thus, by means of that lemma, one can propose for a classical (linear) solution:

$$\frac{6.55\text{m}}{(17\text{m/s} - 1.55\text{m/s})} = 0.42394822\text{s}$$

$t = 1.00 + 0.42394822 = 1.42394822$s

While our proposal (first solution) offered a total time ($t$) of 1.4079585s, the classical solution assumed that A catches B, offered us a time ($t$) of 1.42394822s.

One of the main problems for Zeno's AR paradox and for the classical solution *prima facie*, is the assumption of infinitesimal atomism applied on both, space and movement. Above of all, Zeno's AR paradox assumes that the lynx (Achilles in Aristotle' and seemingly in the original) will stop as it reaches the limit of every segment covered; it was set in a 2 dimensional plane; and both were set at uniform speeds (which magnitudes are the mean value of space over time). On the other hand, our proposal considered spacetime as a continuum (GR); that, whether the hunter (A) sets its goal according to its current (starting) speed and beyond the starting position of B, or to the snail's preceding position; it runs in a 3D plane at non-linear (accelerated) speed, or both (4D); it is doing a race within a gravitational field; and the intrinsic features of the animals (diversity), i.e., the intrinsic properties of both animals, these make one or the other faster (inherent causality relevant to the example), and the specificity that would require to be adjusted for every example otherwise: the lynx runs faster due to its features, and its features made it run even faster, being relevant for the example (even though it was set in idealistic conditions, without taking into account real purposes or lynx's volition to catch a snail, and the snail's speed was set faster for the sake of the example, otherwise, the solution will exploit even better elastic and geometric features diversity and differences between the animals "compared" in a race). In Aristotle's, A is Achilles, whose features made Zeno's AR paradox even more assertive due to Achilles mythical features. Should these features be made for the empirical example, Achilles could be perfectly set as A and the lynx as B, if one is assuming A should reach B (an over-dimensioning and hyperbolicity bias that Zeno probably exploited).

[2] Note that in dimensional analysis this solution makes space and time to lose meaning, since it is the accretion of length and an acceleration to overcome the shared flat space (length-surface) with a flat time (acceleration or time-surface). It will be properly discussed in dimensional analysis within the next section.

Finally, to check our proposal, as a mixed solution (as in Fig. 7), for the same setup, but A at 17 m/s, at a distance of 22m from B escaping (in the same direction) at 12.1 m/s, A reaches (22+12.1)m in $34.1/17 = 2.00588\text{s}$ , then B will have advanced $12.1\ \text{m/s} \cdot 2.00588\text{s} = 24.2711\text{m}$ (it implies more than 1 cycle, until B is located at less than 1 second). Then, A would cover the new distance in $24.2711\text{m} / 17\ \text{m/s} = 1.42771\text{s}$ ; in which the animal B will travel $1.42771\text{s} \cdot 12.1\ \text{m/s} = 17{,}275\text{m}$. ; Now, A will cover it in $17{,}275\ \text{m} / 17\text{m/s} = 1.0161\text{s}$; Finally A in a shared-coordinate, assumed to catch B, it will do it in $2.00588 + 1.42771 + 1.0161 + \frac{1.0161-1}{17} = 4{,}45075\text{s}$ .

The outcome by the previous method being: $22/(17 - 12.1) = 4{,}4897\text{s}$.

Therefore, a mixed answer yields an outcome even lighter than the standard, with a reduction due to what seems pendular gravitational energy, of $\approx 4{,}4897 - 4{,}45075 = 0{,}039046$s and a difference to efficiency with regard to that *pendular energy* of: $\left(\frac{4.4897-4.45075}{22}\right) \cdot 100 - \left(\frac{1.42394822-1.4079585}{22}\right) \cdot 100 = 0.105\%$ of increase per meter (difference between the snail and animal B in the last example, noticing that the speed of the snail was set at 1.55m/s). A faster animal B will increase the expected time, while a slower will decrease the time with respect of the expected time to be caught.

The diversity principle considered is tackling on the difference between the animal A and B, in such a way that, whereas A have a geometric structure which stands higher than B, is built as an elastic structure made for motion and a relatively faster pace than B, and its motion is performed by running (a series or sequence of jumps that allows to exploit gravitational potential energy, in a sort of efficient way referred as pendular energy (Cavagna, Willems, & Heglund, 2000), garnering that energy; B on the other hand, is designed and moves (essentially as it is designed) also garnering some gravitational potential energy, but in a wave-like, specifically in an inefficient wave-like form (snail), so it does not exploit gravitational potential energy that much.

In the last setup, and in the other two, it is assumed that the lynx catches the snail and the other (faster) animal B. For the hyperfinite and infinitely divisible space, it remains as a philosophical argument, and, even though the proposal herein described will hint to a solution and a possible explanation, it will still remain within the metaphysical realm. Notwithstanding, it makes the case for a mechanism such as decoherence (Gao, 2013; Penrose, 1996), as a bypath when two idealistic bodies, should they be considered different, are close enough to postulate that there might be a fundamental length; and also as a bypass between gravity and the quantum world, in which gravity upgrades the displacement of animals featuring elastic properties and motion from uniformly paced to accelerated ones. Decoherence (Gao, 2013;

Penrose, 1996) or minimally coupled to gravity interface with a contact energy (Shieh, 2015) seems to be the mechanism by which, for instance A, can “jump” from the hyperfinite and infinitely divisible space (the Zeno’s AR paradox setup potentially related to the Weyl curvature hypothesis) to the final step into catching the snail (added to the development of paws and articulated limbs). To illustrate that with an example, one may set an analogy, by imagining a shared coordinate between A and B in the Zeno’s AR paradox, setting an analogy between the flat space, of any fundamental length if it is one’s desire, and a hole. We proposed a solution based on a mixed-state that includes a pure-state. When they are within the shared-coordinate framework, it is as if they were approaching that hole, the hole meaning "to catch the snail" for the lynx (A), and "escape" for the snail (B). Only one can come about into the hole, with gravity, geometric setup, and elastic properties (and perhaps a little bit of anticipation) helping the lynx within the GR framework in this case (higher probabilities if it were to be regarded from the QM approach), so it is almost deterministic, yet probabilistic, that the lynx will catch the snail, in that case within the computed lapse.

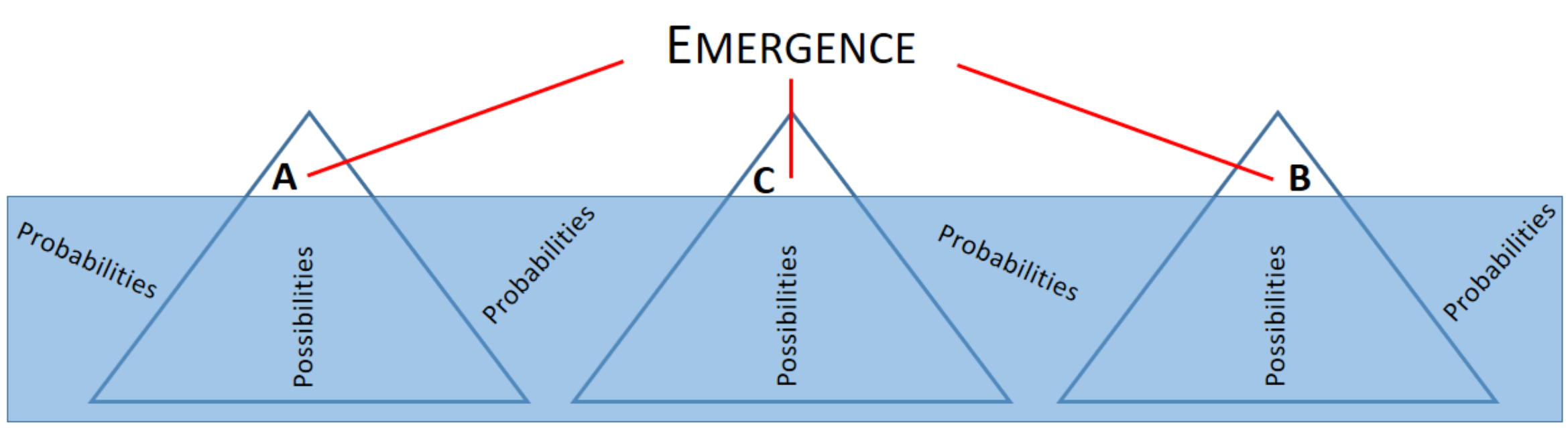

**Figure 7**

The interaction between mass-gravity (boson), geometric and elastic properties of the bodies, and emergent properties and most probable or proximal realities (emergence) will add the reality to the tops of the icebergs.

Usually, the former would become actuality due to supervenience, and the latter from *beables* and QM properties (other effects and eigenstates).

By supervenience and eigenselection, it would produce emergence of one/many of the many reality/ies in $T_0$, ending, potentially, in the decoherence by gravity-mass and the previous bound-states, yielding to a shared reality (past), which can be measured in $t_0$-$t_1$-$t_2$-$t_3$...

As shown in Fig. 7, not only A and B, but also A-B shared coordinate integrate the system. One may argue that the system consists of A and B, plus another sector which could include another animal/s and elements to add up to the complexity, in a time-like or a space-time coordinate. However the case, if it is postulated, that model would pertain and make the case for a potential unified framework with multiple realities comprising the QM world at the emergence level, before one shared-reality (the past), is resolved for the exact very same spacetime coordinate (for all the tops) with the utmost certainty for the eigenstate that "entered" the "hole", due to supervenience (Bache, 2024a), something that have been proposed as the collapse of the wave-function (Ghirardi et al., 1988; Penrose, 1996; 2000; Penrose, 2005, pp. 527-530), by decoherence-like processes, or other processes.

Once a realistic, yet idealistic, example is set, one may postulate that a snail at 0.009 m/s, at 22m, distance from its hunter, a lynx in this case, at 17 m/s, it would have been reached in $(22 + 0.009)/17 = 1.29464706\text{s}$ ; and, to respond to Zeno's AR paradox, the lynx would catch the snail (including the approximation space) at $1.29464706 + ((1.29464706 - 1)/17) = 1.311979239\text{s}$ . One may check that, a solution with the time within the shared space-time coordinate calculated with A's speed over B's speed offers a reduction to slower animals (less efficient, elastic-wise), and an increased ratio for faster animals (more efficient, elastic-wise). Thus, that way allows the introduction for a realistic behavior: faster animals being hunted will feint more, and their momentum will be greater to do it in a more effective way as far as dodging racing hunters is concerned. Both solutions are limited, nonetheless, by the natural setting, group-deception and obstacles, dodging capacity of individual preys, and physiological parameters related to distances, as well as psychological parameters such as the central-controller hypothesis (Noakes, 1997; 2001; St. Clair Gibson, Lambert, & Noakes, 2001), the latter being suggested to be possible to be overcome and overridden by humans (Esteve-Lanao et al., 2008).

# 6 Inertial Torque. A "Bits from It" Explanation and Answer to Zeno's Paradox of Motion

First of all, whether a cheetah or a lynx as A, their features make a case for an accelerated instead of a uniformly paced speed within the race between A and the snail (B). In our proposal, not only does the lynx catch the snail, but it is predicted to do it faster than the prediction given by a linear (classical) setup. As this last one obtains its magnitudes from mean speed (displacement over time), it opens the result to Jensen's inequalities argument (Denny, 2017; Jensen, 1906; Ruel & Ayres, 1999). Although both are relevant for the example, the former is key for an empirical explanation. Since their relative motion within the gravitational field, taking into account the principle of fundamental interaction (Bache, 2024a), it integrates gravity (acceleration) even at a linear displacement at constant speed, wherein that gravitational field, both animals exploit gravity at $T_0$ due to their elastic elements, yet the one that exploits it better not only would reach the higher speed, but it also would exploit *it* better for their relative motion, being *it* the interaction between the potential energy of gravity and the geometric (and biomechanical) construction of every animal.

The empirical explanation is that gravity ($g$) is a one-dimensional pressure, that acting within a gravitational field, exerts its pressure on bodies confining mass, exerting or becoming a force, due to its interaction with the mass confined in the bodies. Thus, one may regard it from a dimensional analysis as:

$$g = m/s^2 \quad \text{or} \quad g = \frac{m}{s^2}$$

which physical meaning can be teased out of the equation, as $g$ has a magnitude of length $m^1$ that can be unravelled especially from the observation that if time vanishes, the unit that prevails is that of length ($m^1$), and within the mathematical framework, pertaining to the empirical evidence, it results in the interpretation that gravity ($g$) would be conservative. (For the non-standard analysis it would be $g = \frac{m}{1}$ but it would lose its physical meaning for gravity $g$ being conservative). Another observation hinted from the dimensional analysis is $g$ as a one-dimensional metric magnitude distributed over a temporal surface before interacting with mass).

Regarding pressure and the equivalence principle, one may also set that:

$$p = \frac{F}{m^2}$$

where $p$ is pressure, $F$ is force, and $m^2$ a surface.

Let $F$ be weight P, $\mathrm{P} = \mathrm{M} \cdot g$ where P is weight, M mass and $g$ gravity in $m/s^2$

then,

$$p = \frac{\mathrm{P}}{m^2}$$

Let any surface under the weight (P) vanish, $m^2$ will therefore vanish, then the pressure ($p$) will show the equivalence principle, in action, the one-dimensional action of gravity interacting with mass, i.e. (P) and the conservation of gravity (m) as an acceleration ($m/s^2$).

In regard to our lynx, it means that gravity ($g$) even though being negative in relation to its effort and its goal, it is the interaction between gravity and the mass (a boson), and the elastic properties of the lynx's structure, that allows to increase the speed, faster than expected by Zeno's AR paradox. It is also a claim that, it is, perhaps, decoherence (Penrose, 1979, pp. 583-586; 1996; Gao, 2013), what allows in this case for the lynx to catch the snail, and finally overcome Zeno's AR paradox.

From biomechanics, it is also a reminder that, the lynx has (developed) limbs (and paws, and pointy claws, i.e. $m^1$ or close to it), which allows it to catch its preys, and it moves exploiting the gravitational potential energy, and can also use its own speed as momentum for its limbs, toes, and claws, when entering the hyperfinite (or infinitesimally divisible) area within Zeno's AR paradox setup (and, usually, in natural settings).

Indeed, it is due to gravity that the lynx catches that speeding snail, and probably Zeno's AR paradox the reason why animals like these also developed limbs.

# Conclusion

In this paper, a possible solution to the Zeno's paradox have been offered. It was also checked against classical mechanics (uniformly paced speed solutions). Although there is some disagreement, one can observe that our proposal offered an interpretation from the empirical point of view, which can be tested against *gedanken* experiments, such as the herein proposed. Whether one decides to solve the example with classical mechanics, the standard view, or our proposal, all of them allow to solve Zeno's AR paradox, and refute the arguments offered by Zeno to defend Parmenides' metaphysical doctrine.

On the other hand, we pointed out and highlighted works and arguments in favor of the rejection of logical atomism, and in favor of dichotomy and dualism. Both have been found, and is proposed, as a requirement for the solution of Zeno's AR paradox, and, possibly, regarding the origin of the universe.

The physical meaning of gravity ($g$) from a dimensional analysis, is that gravity is a one-dimensional pressure, that, while interacting with bodies confining mass within a gravitational field, once any surface under a confining-mass body vanishes, it shows how it is exerting its pressure over these bodies, and its conservative nature becomes evident:

Let the surface ($m^2$) under the object within a gravitational field disappear, then the object would fall, as $p = \frac{F}{m^2}$ where p is pressure, F is a force, and $m^2$ a surface in contact.

$p = (\mu \cdot a)/m^2$ , $p = (\mu \cdot g)/m^2$, $p = \frac{\mathrm{P}}{\mathrm{m}^2}$,

where p is pressure, μ a mass, a an acceleration, g gravity in $m/s^2$, $m^2$ a surface or a length squared, and P stands for weight (μ · g), so, if $m^2$ disappears it means (physical meaning) that $p = \frac{\mathrm{P}}{0}$ without the constraints of mathematics and regarding only the physical meaning:

$$p = \mathrm{P}$$

where p is pressure, and P stands for weight (μ · g), then, if one makes the mass disappear and sets time as 0 as a time-reversal from a gravitational potential,

$p = \mu \cdot g$ , $p = m/s^2$ , $p = (\frac{m}{0})$ , $p = m/s^2$ , $p = m$

one may clear μ and time (s) up, or make it zero, restraining its effect to the physical meaning, with no constraints from mathematics, or keep it acquiring the physical meaning of being conservative.

Indeed, it is possibly due to gravity that the lynx catches the speeding snail, and probably Zeno's AR paradox the reason why animals like these also developed limbs. Our proposal took into account animals' features, integrating inherent movement operators and properties relevant for movement, diversity between animals, and their geometric construction, keeping the physical meaning and the outcome in agreement with theory, over the positivist point of view; and offered an empirical interpretation.

It allowed to explore a possible link between empirical interpretation, gravity's physical meaning and underlying nature, and decoherence as a potential mechanism to overcome paradoxes such as Zeno's AR paradox, and, perhaps, offered hints for a deeper understanding within the foundations of physics and for approaches beyond curtailing, overly axiom-based (unaware idealistic yet restrictive), and imposing mathematical models and approaches, that detach the field (of physics) from the evidence and the experimental (empirical) and constructive approaches.

## Declarations

The authors hereby declare that there are neither funding nor competing interest to declare.

Both authors have contributed to this work and approved it for publication.

## Ethical Approval

This article does not contain any studies with human participants performed by any of the authors.

## Informed Consent

This article does not contain any studies with human participants performed by any of the authors.

## Data Availability

Data sharing is not applicable to this research as no data were generated or analysed.